\documentclass[a4paper, 10pt, conference]{IEEEtran}

\ifCLASSINFOpdf
\else
\fi
  \usepackage[caption=false,font=footnotesize]{subfig}

\usepackage[cmex10]{amsmath}

\usepackage{color}
\usepackage{fixltx2e}
\usepackage{enumitem}

\usepackage{amsfonts}

\usepackage{amssymb,hhline}
\usepackage{amsthm}

\usepackage[ruled,vlined,linesnumbered,lined,boxed]{algorithm2e}
\usepackage{graphicx}
\usepackage{graphicx,subfig}
\usepackage{multicol}
\usepackage{booktabs}
\usepackage{multirow}
\usepackage{tabularx}
\usepackage{dblfloatfix}    

\usepackage{array}
\newcolumntype{m}[1]{>{\centering\arraybackslash}p{#1}}

\usepackage{bm}

\newcommand{\bom}[1]{\boldsymbol{#1}}
\newcommand{\bo}[1]{\mathbf{#1}}

\newcommand{\x}{\mb}
\newcommand{\supp}{\mathrm{supp}}

\newcommand{\y}{\bo y}

\newcommand{\X}{\bo X}   
\newcommand{\R}{\mathbb{R}}    

\newcommand{\mdim}{m}             
\newcommand{\ndim}{n}             
\newcommand{\pdim}{p}             

\newcommand{\paino}{\sl}

\renewcommand{\a}{\bo a}

\newcommand{\mb}{\bo x}    

\newcommand{\beq}{\begin{equation}}
\newcommand{\eeq}{\end{equation}}
\newcommand{\bmat}{\begin{pmatrix}}
\newcommand{\emat}{\end{pmatrix}}
 \newcommand{\beqa}{\begin{eqnarray}}
\newcommand{\eeqa}{\end{eqnarray}}

\newcommand{\eps}{\bom \varepsilon} 
\newcommand{\Sig}{\bom \Sigma} 
\newcommand{\be}{\bom \beta}
\newcommand{\al}{\alpha} 
\newcommand{\sig}{\sigma}  
\newcommand{\lam}{\lambda}  
\newcommand{\gam}{\gamma}

\newcommand{\lamh}{\lam} 

\newcommand{\beh}{\hat{\be}}  
\newcommand{\bet}{\be^{\ast}}  

\newcommand{\XA}{\X_{\mathcal A}} 
\newcommand{\setA}{\mathcal A}

\usepackage{cite}

\makeatletter
\newcommand*\titleheader[1]{\gdef\@titleheader{#1}}
\AtBeginDocument{%
	\let\st@red@title\@title%
	\def\@title{%
		\bgroup\normalfont\small\centering\@titleheader\par\egroup
		\vskip0.5em\st@red@title}
}
\makeatother
\graphicspath{ {Results/} }

\title{Pathwise Least Angle Regression  and \\a Significance Test for the Elastic Net}  

\titleheader{This paper is accepted to appear in the proceedings of the 25th European Signal Processing Conference (EUSIPCO), 2017}

\author{\IEEEauthorblockN{Muhammad Naveed Tabassum and Esa Ollila}
	\IEEEauthorblockA{Aalto University, Dept. of Signal Processing and Acoustics, P.O. Box 15400, FI-00076 Aalto, Finland} 
	Email: \{muhammad.tabassum, esa.ollila\}@aalto.fi }

\begin{document}
%

\maketitle

\begin{abstract}
Least angle regression (LARS)  by Efron et al. (2004) is a novel method for constructing the piece-wise linear path of Lasso solutions. For several years, it remained also as the de facto method for computing the Lasso solution before more sophisticated optimization algorithms preceded it. LARS method has recently again increased its popularity due to its ability to find the values of the penalty parameters, called \emph{knots}, at which a new parameter enters the active set of non-zero coefficients. Significance test for the Lasso by Lockhart et al. (2014), for example, requires solving the knots via the LARS algorithm.  Elastic net (EN), on the other hand, is a highly popular extension of Lasso that uses a linear combination of Lasso and ridge regression penalties. 
In this  paper, we propose a new novel algorithm, called  pathwise (PW-)LARS-EN, that is able to compute the EN knots over a grid of EN tuning parameter $\al$ values. 
The developed PW-LARS-EN algorithm decreases the  EN tuning parameter and exploits the previously found knot values and the original LARS algorithm.
A covariance test statistic for the Lasso is then generalized to the EN  for testing the significance of the predictors. Our simulation studies validate the fact that the test statistic has an asymptotic $\mathrm{Exp}(1)$ distribution. 

\end{abstract}

%
\IEEEpeerreviewmaketitle

\section{Introduction}

In this paper, we consider a linear model, where the $n$-vector $\y  \in\mathbb{R}^{n}$ of observations is modeled as 
\beq\label{eq:linear} 
\y = \X \be + \eps,
\eeq
where $\X\in\mathbb{R}^{n\times p}$ is a known predictor matrix, $\be \in\mathbb{R}^{p}$ is the  unknown vector of regression coefficients  and 
$\eps \in\mathbb{R}^{n}$ is the noise vector. 
For ease of exposition, we consider the centered linear model (i.e. we assume that the intercept is equal to zero). 

Elastic net (EN) of \cite{zou2005reg} is a superset of the popular Lasso (Least absolute shrinkage and selection operator) \cite{lasso:1996} that is also termed as basis pursuit denoising (BPDN) in the  literature. 
EN has recently been employed, for example, in a single snapshot DoA (direction-of-arrival) finding application in \cite{tabassum2016single}. 
The EN estimator is defined as the solution of the following penalized residual sum of squares (RSS) optimization problem,
\beq\label{eq:penEN} 
\hat \be(\lam,\al) = \underset{\be \in \mathbb{R}^\pdim}{\arg \min}  
 \ \frac  1 2  \|\y - \X \be \|_2^2   + \lam P_\al\bigl(\be \bigr) 
\eeq 
where $\lam \geq 0$ is the {\paino EN penalty parameter} and the EN penalty term $P_\al\bigl(\be \bigr)$,  defined as  
\[
P_\al\bigl(\be \bigr) = \al\|\be\|_1 + \cfrac{(1-\al)}{2} \| \be \|_2^2 ,  
\]
is a convex  combination of $\ell_1$-norm and $\ell_2$-norm penalties of the Lasso and ridge regression. 
The  {\paino EN tuning parameter} $\al \in [0,1]$, which is chosen by the user,  determines the mix between ridge regression and the Lasso. 
The Lasso is obtained for $\al=1$, and will be denoted shortly as $\hat \be(\lam)$ instead of $\hat \be(\lam,1)$. 
The EN penalty has singularities at the vertexes like Lasso, which is a necessary property for sparse estimation. It also has strictly convex edges that then help in selecting variables as a group, which is a useful property when high correlations exists between predictors.  


For  a given fixed EN tuning parameter $\al$, the path of solutions $\hat \be(\lam,\al)$ indexed by $\lam$ are also piece-wise linear as in the case of  Lasso ($\al=1$). 
It is then of interest to find the values of the penalty parameters, $\lam_0,\lam_1,\ldots, \lam_K$,  called the {\paino knots}, at which a new parameter enters the active set of non-zero coefficients.
In the special case of Lasso, this can be achieved via the famous {\paino least angle regression (LARS) algorithm} \cite{efron2004LAR}.
In this paper, we propose a pathwise (PW-)LARS-EN algorithm that computes the knots of EN over a grid of $\al$ values. 
The developed PW-LARS-EN algorithm starts with $\al=1$ and computes the knots via the LARS algorithm. We then decrease the EN tuning parameter and exploit the knowledge of  the previously found values of the knots and the original LARS algorithm (using an augmented form of the EN regularization problem) to compute the knots of EN at current $\al$ in the grid.   
We decrease $\al$ again, and repeat the procedure. In this way we can efficiently compute the solutions over a grid of $\al$ values. 


Recently, in  \cite{lockhart2014CovTest}, a  covariance test statistic $T_k$ was proposed for testing the significance of predictors in the context of the Lasso. 
The covariance test statistic requires solving the knots via the LARS algorithm.   In this paper, we generalize the covariance test statistic  for the EN, which we denote as $T_k(\al)$. This test statistic then requires computing the knots of EN solution for a given fixed $\al \in [1,0)$.    
It was postulated in  \cite[Sec. 8]{lockhart2014CovTest} that covariance test statistic for the EN also follows standard exponential distribution,  $\mathrm{Exp}(1)$, when the null hypothesis that 
all signal variables are in the model holds true. The authors, in \cite{lockhart2014CovTest}, proved this fact in the orthonormal case, i.e.,  when  $\X^\top \X = \bo I_{\pdim}$ and $\ndim=\pdim$. 
Thanks to the proposed PW-LARS-EN algorithm, we are able to compute the empirical distribution of  $T_k(\al)$  also in the general non-orthogonal case. 
Our simulation studies then confirm that the $\mathrm{Exp}(1)$ approximation for the covariance test statistic $T_k(\al)$ is valid in the general case as well.


The paper is organized as follows. In Section~\ref{sec:LARS-EN}, we describe and derive  the pathwise LARS-EN algorithm which computes the knots of EN estimators over a grid 
of EN tuning parameter values. The covariance test statistic for the EN is described in Section~\ref{sec:covtest}, and Section~\ref{sec:simul} presents the simulation study. A real data example is given in Section~\ref{sec:data} while Section~\ref{sec:concl} concludes the paper.

{\it Notations}: 
Uppercase boldface letters are  used for matrices and  lowercase   boldface  letters for  vectors. 
The vector space  $\mathbb{R}^{\ndim}$ is equipped with the usual inner product, $\langle \a, \bo b \rangle = \a^\top \bo b$,  where $(\cdot)^\top$ denotes the transpose. This induces the conventional $\ell_2$-norm $\| \a \|_2 = \sqrt{\a^\top \a}$.   The $\ell_1$-norm is defined as $\| \a\|_1 = \sum_{i=1}^\ndim |a_i |$, where $| a | $ denotes the absolute value  of $a\in \R$.  The $\ell_0$-norm of a vector is defined as $\| \a\|_0 = \#(i|a_i\neq 0)$, which is equal to the total number of non-zero elements in it. 
The support of $\a \in \R^\pdim$ is the index set of its non-zero elements, i.e., $\supp(\a)= \{ j \in \{ 1, \ldots, \pdim\} : a_j \neq 0 \} $.

\section{Pathwise LARS-EN Algorithm} \label{sec:LARS-EN}

It is well-known that the paths of solutions of  $\hat \be(\lam)$ as a function of the regularization parameter  $\lam$ are piece-wise linear in each coefficient.  
The values of $\lam$ at which a new predictor variable becomes active (and hence a change in the slope occurs) or leaves the active set are referred to as  {\paino knots} and  are denoted as  $\lam_0> \lam_1> \ \ldots > \lam_K$. 
The values of the knots are not fixed, but depend on the data $(\y, \X)$. 
LARS-Lasso  delivers the entire solution path as a function of the regularization parameter $\lam$. For a detailed discussion of LARS-Lasso algorithm, we refer the reader to \cite{ efron2004LAR} or \cite[Sect.~ 5.6]{hastie2015stat}. 
 In essence, the LARS-Lasso  finds the knots. The knowledge of the knots then allows to solve the whole coefficient paths as the 
coefficients either increase or decrease in the interval $(\lam_k,\lam_{k+1})$. 

It is also well-known that the EN estimator $\beh(\lam,\al)$ has piece-wise linear  solution paths for a given fixed $ \al \in [1,0)$. 
We now develop the PW-LARS-EN algorithm for finding the knots $\lam_0(\al),\lam_1(\al), \ldots, \lam_K(\al)$ of the EN solution in  \eqref{eq:penEN} for fixed $\al$. 
Let $\lamh_0(\al)$ denotes the smallest value of $\lam$ such that all estimated coefficients are zero, i.e., $\beh(\lamh_0(\al),\al)=\bo 0$.  This value is easily shown to be 
\[ 
\lamh_0(\al) =  \frac{ \mathrm{max}_j    | \langle \x_j, \y \rangle | } {\al }, 
\]
where maximum is over $j \in \{ 1,\dots,p \}$. 
Let  $\setA(\lam, \al)$ denotes the {\paino active set} at  $\lam < \lamh_0(\al)$,  i.e., the index set of predictors with nonzero coefficients values:
\[
\setA(\lam, \al) = \mathrm{supp}\{\beh(\lam,\al)\}
\] 
Thus $ [\beh(\lam,\al) ]_j \neq 0$ for all $j \in \setA(\lam, \al)$. Each knot is a  border value, after which a change in the set of active predictors occurs. The active set at the knot $\lam_k(\al)$ is denoted by  $\setA_{k}(\al) =  \setA(\lam_k,\al)$. 
The active set $\setA_1(\al)$ thus contains a single index $\setA_1(\al) = \{ j_1 \}$, where $j_1$ is the predictor that becomes active first and is known to be $j_1 = \ \arg \max_j | \langle \x_j , \y \rangle | / \al $. By definition of the knots, one has that $\setA_k(\al) = \{\mathrm{supp}\{\beh(\lam,\al)\} $ for all $\lam_{k-1} (\al) < \lam  \leq \lam_k(\al)$ and $\setA_{k}(\al) \neq \setA_{k+1}(\al)$ for all $k=0,1,\ldots,K-1$. We denote by $\XA$ the  matrix $\X$ restricted to the columns of the active set $\setA$.

The PW-LARS-EN algorithm finds the knots  $\lam_0(\al)> \lam_1(\al)> \lam_2(\al) > \ldots > \lam_K(\al)$ for given fixed $\al$ in a grid
\[
[\al] = \{ \al_{i} \in [1, 0) \ :  \   1 = \al_{1} < \al_2 <   \cdots < \al_{\mdim} <0  \}.
\] 
The success of our propposed algorithmic scheme is due to the usage of a dense grid, and therefore we use $[\al] =\{ 1, 0.99, 0.98, \ldots \}$ as the default grid. This grid is used  in the simulations and data analysis examples of the paper. It should be noted that $\al_m$ is the smallest value of EN parameter $\al$ for which the user wishes to obtain the EN solution paths. Often the regime $[0.5, 0)$ may not be very interesting as those EN solutions are closer to the Ridge regression estimator ($\al=0$) than the Lasso estimator ($\al=1$). In our examples,  $\al_m$ is often chosen as  $\al_m=0.5$.    
The PW-LARS-EN algorithm first finds the knots for the Lasso (i.e. the case of $\al=1$) using the LARS-Lasso.  We then decrease $\al$ and  consider the first value $\al_1$  in the $\al$-grid, and exploit the augmented form in (\ref{eq:augEN}) for subsequent values of $\al$.

\graphicspath{ {Results/} }
\begin{figure*}[!b]
\centering
	\subfloat[]{\includegraphics[width=0.32\textwidth]{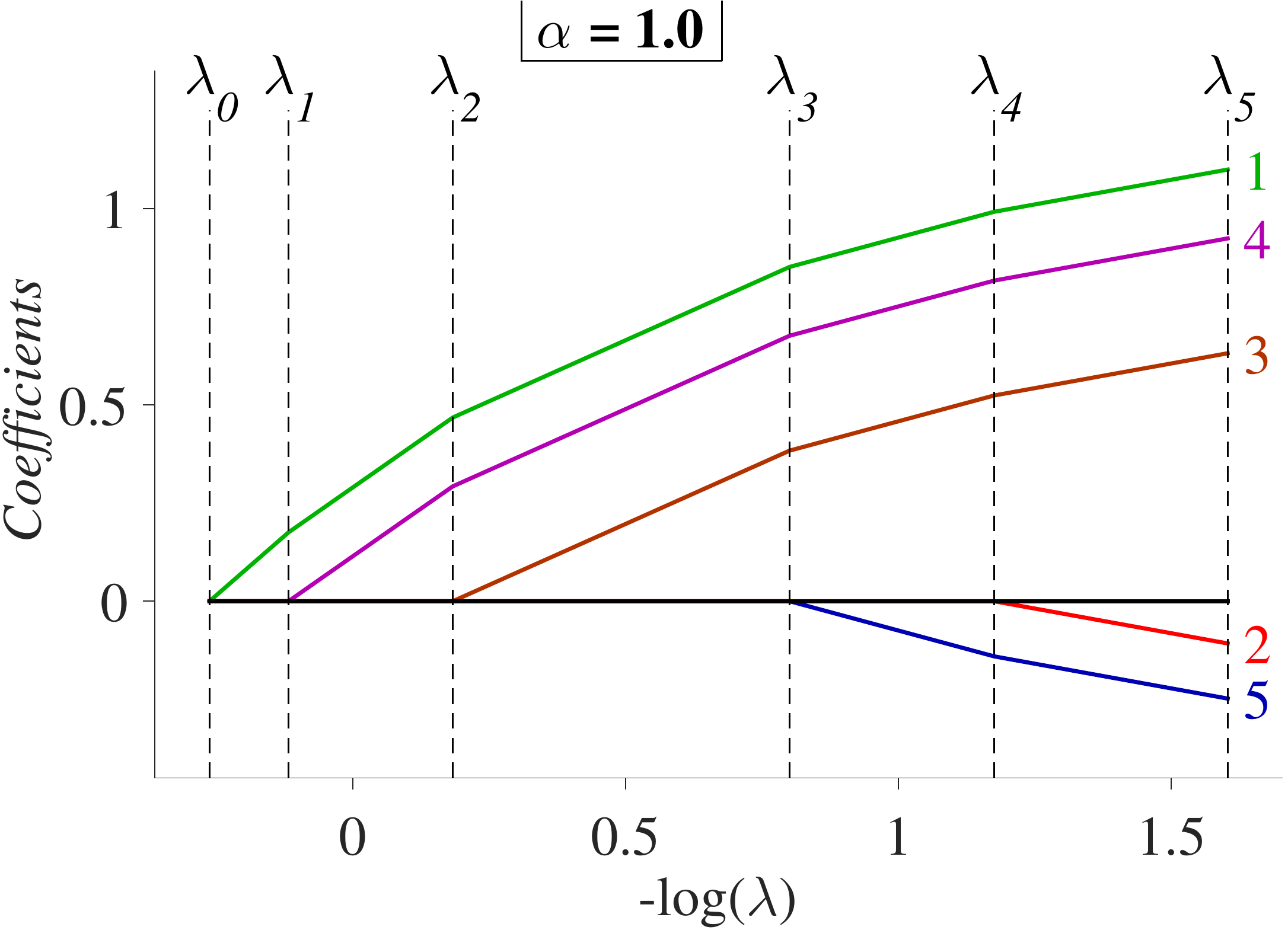}}\quad
	\subfloat[]{\includegraphics[width=0.32\textwidth]{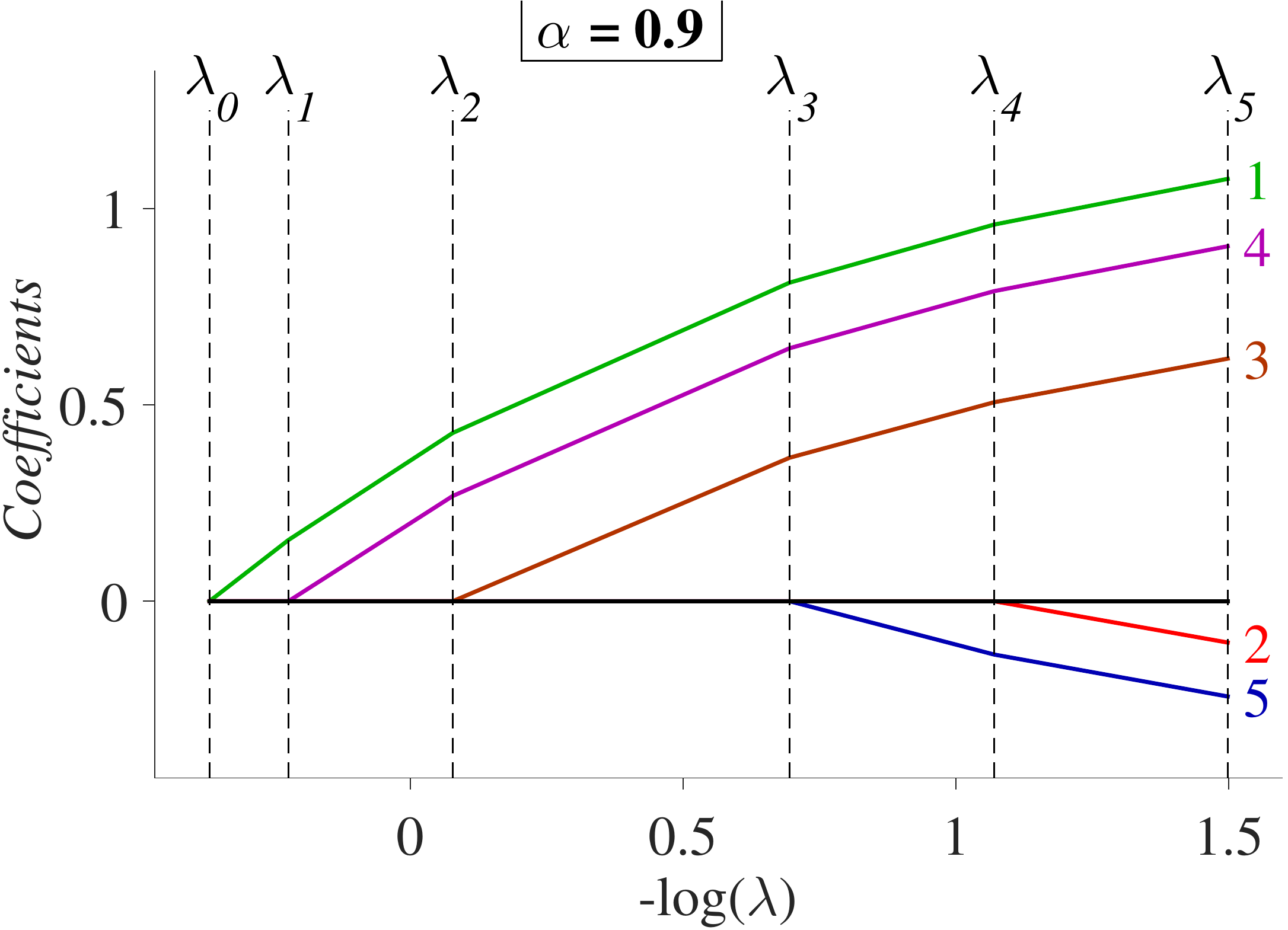}}\quad
	\subfloat[]{\includegraphics[width=0.31\textwidth]{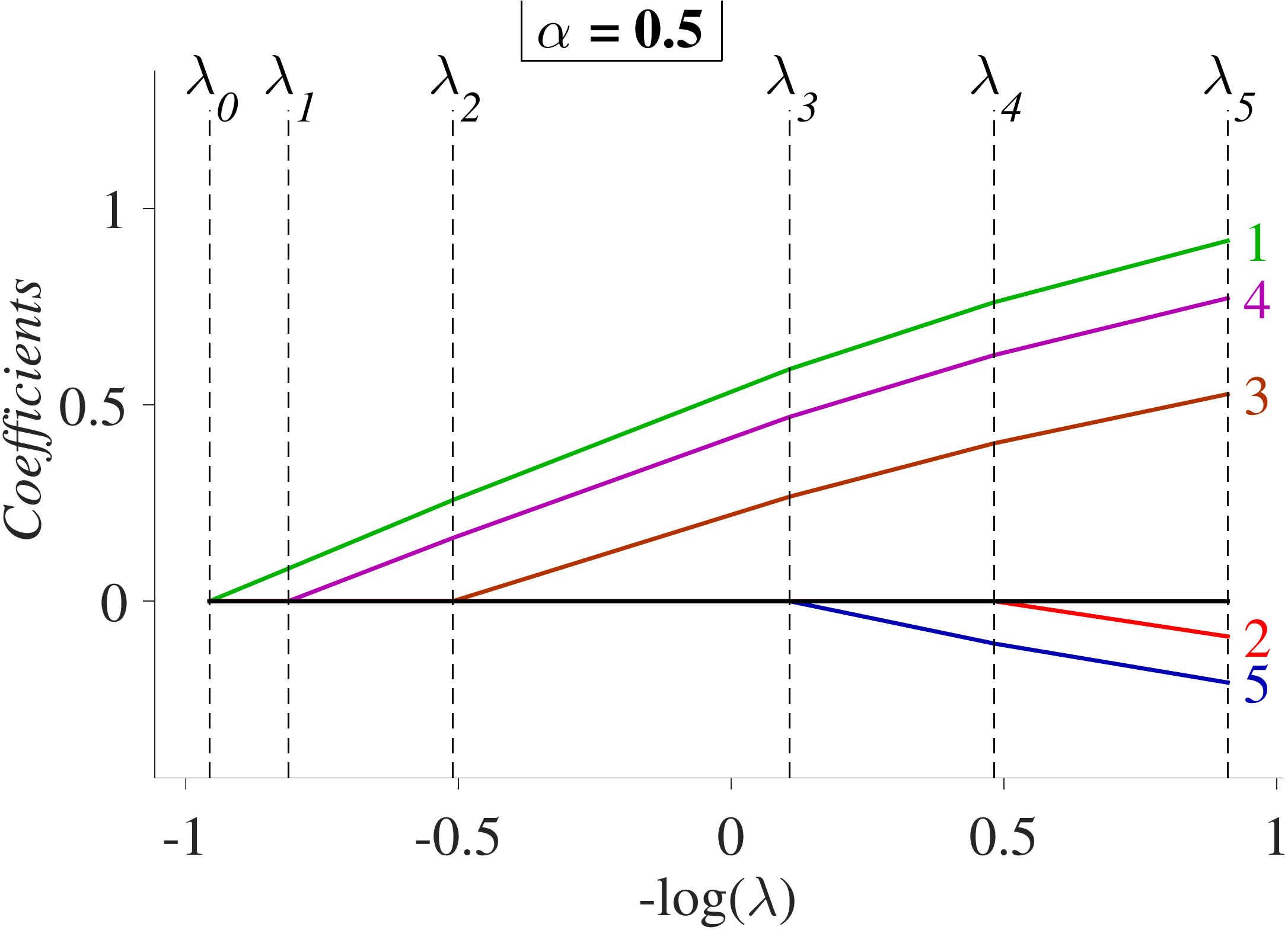}}\hfill	
\caption{Elastic net coefficient profiles for three different values of $\al$ and the found knots via PW-LARS-EN algorithm on an simulated data ($n=100$,  $p=10$), where the true coefficient vector $\be^*$ has 5 non-zeros.  Note that the left panel corresponds to Lasso ($\al=1$).  } 
\label{fig:CPplots}
\end{figure*}

Let $\mbox{LARS}\big(\y, \X\big) $ denotes the LARS-Lasso algorithm that computes the knots $\{\lam_k\}_{k=0}^K$ for the Lasso along with corresponding solutions at the knots $\{\hat \be(\lam_k)\}_{k=0}^K$. Let  
\[ 
\{ \lam_k, \beh(\lam_k) \} = \mathrm{LARS}\big(\y, \X) \big|_k  
\] 
denotes the case that the $k^{th}$ knot and the respective solution is extracted from a  sequence of the  knot-solution pairs found by the LARS-Lasso algorithm. 
Next note that we can write the EN objective function in  {\paino augmented form} 
as follows:   
\beq\label{eq:augEN} 
\frac 1 2 \| \y - \X \be \|_2^2 + \lam P_\al(\be) = 
 \ \frac  1 2 \| \y_a   - \X_a(\eta)  \be \:\|_2^2   + \gam \big\| \be \big\|_1 
\eeq 
where  
\beq \label{eq:gam_eta} 
	\gam  = \lam \al \qquad \text{and} \qquad \eta = \lam (1-\al), 
\eeq 
are new parameterizations of the tuning and shrinkage parameter pair $(\al,\lam)$,  and  
\[ 
\y_a = \bmat \y \\ \bo 0 \emat \qquad \text{and} \qquad  \X_a(\eta)  =  \bmat \X \\ \sqrt{\eta} \, \bo I_\pdim  \emat 
 \] 
are  the {\paino augmented}   forms of the response vector $\y$ and the predictor matrix $\X$, respectively. Note that \eqref{eq:augEN} resembles the Lasso objective function with $\y_a \in \R^{\ndim + \pdim}$ and that
 $\X_a(\eta)$ is an $(\ndim + \pdim) \times \pdim$ matrix.
Next consider the case of finding  the knot  $\lam_k(\al_i)$ for the $i^{th}$ value,  $\al_i$, in the grid $[\al]$.   Equivalently stated, we wish to find the pair 
\[ 
\gamma_k = \lam_k(\al_i)  \cdot \al_i \qquad \text{and} \qquad  \eta_k = \lam_k(\al_i) \cdot (1-\al_i)
\]
using the alternative parametrization given in \eqref{eq:gam_eta}.   Obviously, the knowledge of either $\gamma_k$ or $\eta_k$ for any fixed $\al_i$ would allow us to solve for 
$\lam_k(\al_i)$.   In our algorithm, we  exploit the following property:  If we would know  $\eta_k$,  then \eqref{eq:augEN} is simply a Lasso problem and due to equivalence \eqref{eq:augEN},  
$\gamma_k$ is the $k^{th}$ knot of  the Lasso problem for the augmented data $(\y_a,\X_a(\eta))$.  Thus we can find $\gamma_k$ and $\beh(\lam_k,\al_i)$  via the LARS-Lasso algorithm as  
\[
\{ \gamma_k, \beh(\lam_k,\al_i) \}  = \mathrm{LARS}(\y_a, \X_a(\eta_k) ) \big|_k .  
\]
The knot $\lam_k(\al_i)$ is then found simply by scaling the found knot $\gam_k$ as $\lam_k(\al_i) = \gam_k / \al_i$. 
Naturally, we do not know $\eta_k$, but fortunately a good approximation can be found  using the $k^{th}$ knot  that was 
found  previously (for $\al_{i-1}$).  That is, we use  
 \[
\tilde \eta_k  = \lamh_k(\al_{i-1}) \cdot (1-\al_i) \approx \eta_k. 
 \] 
Algorithm~\ref{algo:larsen} outlines the pseudo code for this procedure for any given value of $k$. It presumes that predictors are standardized ($ \| \x_j\|^2 = 1$).  

\setlength{\textfloatsep}{2pt}
\begin{algorithm}[!h]
\DontPrintSemicolon
 \caption{PW-LARS-EN algorithm.}\label{algo:larsen}
\SetKwInOut{Input}{input}\SetKwInOut{Output}{output}
\BlankLine
\Input{$\quad \y\in\mathbb{R}^{n}$, $\X\in\mathbb{R}^{n\times p}$, $[\al] \in\mathbb{R}^{\mdim}$ (recall $\al_1=1$). 
}
\BlankLine
\Output{\quad $\{ \lamh_k(\al_i), \beh(\lam_k,\al_i) \}_{k=1,i=1}^{K,m}$ \, .} 
\BlankLine
\BlankLine
$ \{ \lamh_k(\al_1), \beh(\lam_k,\al_1) \big\}_{k=1}^K = \text{LARS}\big( \y,\X \big) $

\BlankLine
\BlankLine
   \For{$i=2$ \KwTo $\mdim$}{
      \For{$k=2$ \KwTo $K$}{

\BlankLine
 $ \tilde \eta_k  =  \lamh_k(\al_{i-1}) \cdot (1-\al_i) $ 
  \BlankLine
$\big\{ \gam_k , \beh(\lamh_{k},\al_i) \big\} = \mathrm{LARS}\big( \y_a, \X_a(\tilde \eta_k) ) \big|_k $
\BlankLine
$\lamh_k(\al_{i}) = \gam_k/\al_i \quad$
}
}
\vspace*{-1pt}
\end{algorithm}

Let us illustrate the Algorithm on an simulated data set. 
We calculate the EN solution paths for simulated data from $\y \sim \mathcal N_\ndim(\X \bet , \bo I)$, where $n=100$, $\X\in\mathbb{R}^{100 \times 10}$ is an orthogonal predictor matrix, $\bet \in \R^{10}$ contains $k= 5$ non-zero regression coefficients. Fig.~\ref{fig:CPplots} depicts the EN coefficients paths along with the knots (i.e. $\lambda_k(\al)$ values in which a new variable enters the active set) that were found via our PW-LARS-EN algorithm.    
We consider 3 different values of EN tuning parameter $\al = 1$ (Lasso), $\al= 0.9$ and $\al= 0.5$, which are used throughout the paper for reporting the results. 
It can be observed that the proposed PW-LARS-EN algorithm is able to correctly find the EN solution path for each $\al$-value, which is evident from the Fig. \ref{fig:CPplots}. 

\section{Covariance Test for the Elastic Net}\label{sec:covtest}

Recently, in  \cite{lockhart2014CovTest}, a  covariance test statistic was proposed for testing the significance of predictors in the context of the Lasso. 
The covariance test statistic requires solving the knots via the LARS algorithm and it tests the  significance of the predictor that has entered the Lasso solution path in the interval
 $(\lambda_k,\lam_{k+1})$.  Thus we wish to test if the predictor that has entered the active set $\setA_{k+1}$ is just noise or a statistically significant predictor.    
As earlier, let $\lam_k$ and $\mathcal A_k$ denote the $k$-th knot and the active set for the interval $(\lam_{k-1},\lam_{k})$, respectively, for the Lasso.  
The {\paino covariance test statistic} is defined as \cite{lockhart2014CovTest} 
\begin{equation}\label{eq:Tk_lasso} 
T_k =  
\frac{1}{\sigma^2} \Big( \big\langle \y, \X \beh(\lam_{k+1}) \big\rangle - \big\langle \y, \X_{\setA_k} \beh_{\setA_k}(\lam_{k+1}) \big\rangle \Big)
\end{equation}
where $\sigma^2$ is the noise variance. 
Above in (\ref{eq:Tk_lasso}), the notation $\beh_{\setA_k}(\lam_{k+1})$ refers to the Lasso estimate that is computed using the penalty parameter $\lam_{k+1}$ and using just
the variables in $\setA_{k}$. In other words, $T_k$ measures how much of the covariance between $\y$ and obtained fit can be attributed to the predictor which has just entered the model.
Under the null hypothesis that all $k$ signal variables are in $\setA_k$, i.e., $H_0: \setA_{k} \supseteq \setA^*$ where $\setA^*=\supp(\be^*)$ is the underlying true support set,  one has that $T_k \to_d \mathrm{Exp}(1)$ as $\ndim, \pdim \to \infty$.  This result is valid   under the assumption that the noise terms are i.i.d. Gaussian, $\eps \sim \mathcal N_\ndim(\bo 0, \sigma^2 \bo I)$
and under  general conditions on the predictor matrix $\X$. 


The {\paino covariance test statistic} for the  EN estimator that uses  the EN tunining parameter $\al$-value is defined as 
\begin{gather*} 
T_k(\al) =   \\  \frac{  1 + \eta_{k+1} }{\sigma^2}  \Big( \big\langle \y, \X \beh(\lam_{k+1},\al) \big\rangle - \big\langle \y, \X_{\setA_k} \beh_{\setA_k}(\lam_{k+1},\al) \big\rangle \Big) 
\end{gather*}
where $\lam_{k+1}  \equiv \lam_{k+1}(\al)$ denotes the knot of EN estimator for fixed $\al$ and 
$\eta_{k+1} =  \lamh_{k+1} \cdot (1-\al)$. Similarly, $\beh_{\setA_k}(\lam_{k+1},\al)$ refers to an EN estimate for fixed $\al$, which is computed using the $(k+1)^{th}$ knot as the penalty parameter and just the variables in $\setA_{k}\equiv \setA_k (\al)$.

In the simulations that follow, we wish to test validity of the conjecture  that EN covariance test statistic $T_k(\al)$ for any fixed $\al$ also converges asymptotically to standard exponential random variable, i.e.,  $T_k(\al) \to_d \mathrm{Exp}(1)$ as $\ndim,\pdim \to \infty$.
Indeed this was postulated in \cite[Sec. 8]{lockhart2014CovTest} where the authors only showed that $\mathrm{Exp}(1)$-distribution holds true for $T_k(\al)$ when the predictor matrix is orthonormal ($\X^\top \X = \bo I_{\pdim}$).  The authors stated in \cite[Sec. 8.1]{lockhart2014CovTest} that  ``one is tempted to use this approximation beyond the orthogonal setting as well".  Indeed our simulation results reported below confirm that the $\mathrm{Exp}(1)$ approximation for $T_k(\al)$ is valid in the general case as well. 


\section{Simulation Study}\label{sec:simul}

We generate  an  $n\times p$ predictor matrix 
$
\X = \bmat \x_1 &  \x_2 & \ldots & \x_n \emat^\top,
$
where vectors $\x_i \in\mathbb{R}^{p}$,  $i=1,2,\dots,n$, are i.i.d. random vectors distributed as $\x_i \sim \mathcal{N}_p (\bo 0, \Sig(\sigma^2,\rho, \texttt{ST})\, \big)$, 
where $\Sig(\sigma^2,\rho, \texttt{ST}) \in \mathcal{S}^\pdim_{++}$ 
is a $\pdim \times \pdim$ positive definite covariance matrix parametrized by marginal variance $\sigma^2 = \mathrm{var}(x_{i})>0$ and $\rho \in (-1,1)$ which determines the correlation coefficient $\mathrm{corr}(x_{i},x_{j})$ for all $i,j \in \{ 1, \ldots, \pdim\}$.  
The parameter $ \texttt{ST}$  determines the structure which can be 
%
a compound symmetry (CS) covariance structure, 
\[ 
\Sig(\sigma^2,\rho,{\texttt{CS}}) = \sigma^2 \{ (1-\rho) \bo I + \rho \bo 1 \bo 1^\top \} 
\]
(where $\bo 1$ a $\pdim$-vector of $1$s)
or a  first-order autoregressive (AR1) covariance structure in which case  
 the $(i,j)^{th}$ element is 
\[
[\Sig(\sigma^2,\rho,{\texttt{AR1}})]_{ij} = \sigma^2 \rho^{|i-j|} 
\]
for $i, j  \in \{ 1, \ldots, p\}$.


\subsection{Global Null Hypothesis}

Let $\y \sim \mathcal N_\ndim(\X \bet , \bo I)$, where $\bet  = \bo 0 $, i.e., all coefficients are zero and  
the covariance structures for $\x_i$   confirms with  $\Sig_1 = \Sig(1, 0, \texttt{I}) = \bo I_p$, $\Sig_2 = \Sig(1, 0.25, \texttt{CS})$ and $\Sig_3 = \Sig(1, 0.25, \texttt{AR1})$.
For all three cases, we test the global null hypothesis of $\bet=\bom 0$ (i.e. $\| \bet \|_0= k = 0$) via covariance test $T_0(\al)$, 
i.e., we test the significance of the predictor that first enters the active set. 
We generated $n = 100$ observations  and $\pdim$ varied from  $p=10$ to $\pdim = 50$.  
 We compute  the statistic $T_0(\al)$ for  $1000$ data sets simulated from the model above and report the empirical mean, variance, and 95\% quantile. If  $\mathrm{Exp}(1)$ is a good approximation for the empirical distribution of $T_0(\al)$, we expect these figures to be close to 
mean, variance and 95\% quantile of the $\mathrm{Exp}(1)$-distribution. 
The values reported in the Table \ref{table:t1} confirm that the empirical distribution of  $T_0(\al)$  can be well approximated by  $\mathrm{Exp}(1)$ distribution in all the cases of predictor matrix and EN tuning parameter considered ($\al \in \{1.0,0.9,0.5\}$).  

%

\begin{table}[!t]
\setlength\extrarowheight{2pt}
\centering
    \caption{Mean, Variance and $q_{.95}$  Quantile of the Empirical Distribution of $T_0(\al)$ for $\al = (1.0, 0.9, 0.5)$ Over  1000 Simulated Data Sets from the Global Null Model.} \label{table:t1}
\resizebox{0.49\textwidth}{!}{\begin{tabular}{ c | l c | c c c |  c c c }
\cline{4-9}
\multicolumn{3}{ c|}{} &\multicolumn{3}{c|}{$n=100$, $p=10$}  &\multicolumn{3}{c}{$n=100$, $p=50$} \\ 
\cline{2-9} 
  			&	&Exp(1)  	&$1.0$ 	&$0.9$ &$0.5$  & $1.0$ 	&$0.9$ &$0.5$   \\ \hline  

   \multirow{3}{*}{$\Sig_1$}&Mean 		& 1		&0.998 	&0.994 &0.985  &1.014 &1.008 &0.996  \\ 
    						&Var		& 1		&1.269 	&1.255 &1.212  &1.270 &1.248 &1.202  \\ 
    						&$q_{.95}$	& 3.0	&3.200	&3.181 &3.091  &3.044 &3.050 &3.045  \\ 
\hline 

   \multirow{3}{*}{$\Sig_2$}&Mean 		& 1		&0.989 	&0.974 &0.928  &0.993 &0.973 &0.927  \\ 
    						&Var		& 1		&1.423 	&1.408 &1.306  &1.118 &1.078 &0.979  \\ 
    						&$q_{.95}$	& 3.0	&3.118	&3.064 &2.917  &3.051 &2.996 &2.897  \\ 
\hline 

   \multirow{3}{*}{$\Sig_3$}&Mean 		& 1		&0.991 	&0.988 &0.982  &1.009 &0.998 &0.976  \\ 
    						&Var		& 1		&1.619 	&1.618 &1.600  &1.285 &1.258 &1.189  \\ 
    						&$q_{.95}$	& 3.0	&3.311	&3.266 &3.261  &3.241 &3.198 &3.031  \\ 
\hline 

\end{tabular} }
\end{table}

\subsection{General Null Hypothesis}

\begin{figure}[!t]
\centering
	\subfloat[]{\includegraphics[width=0.163\textwidth]{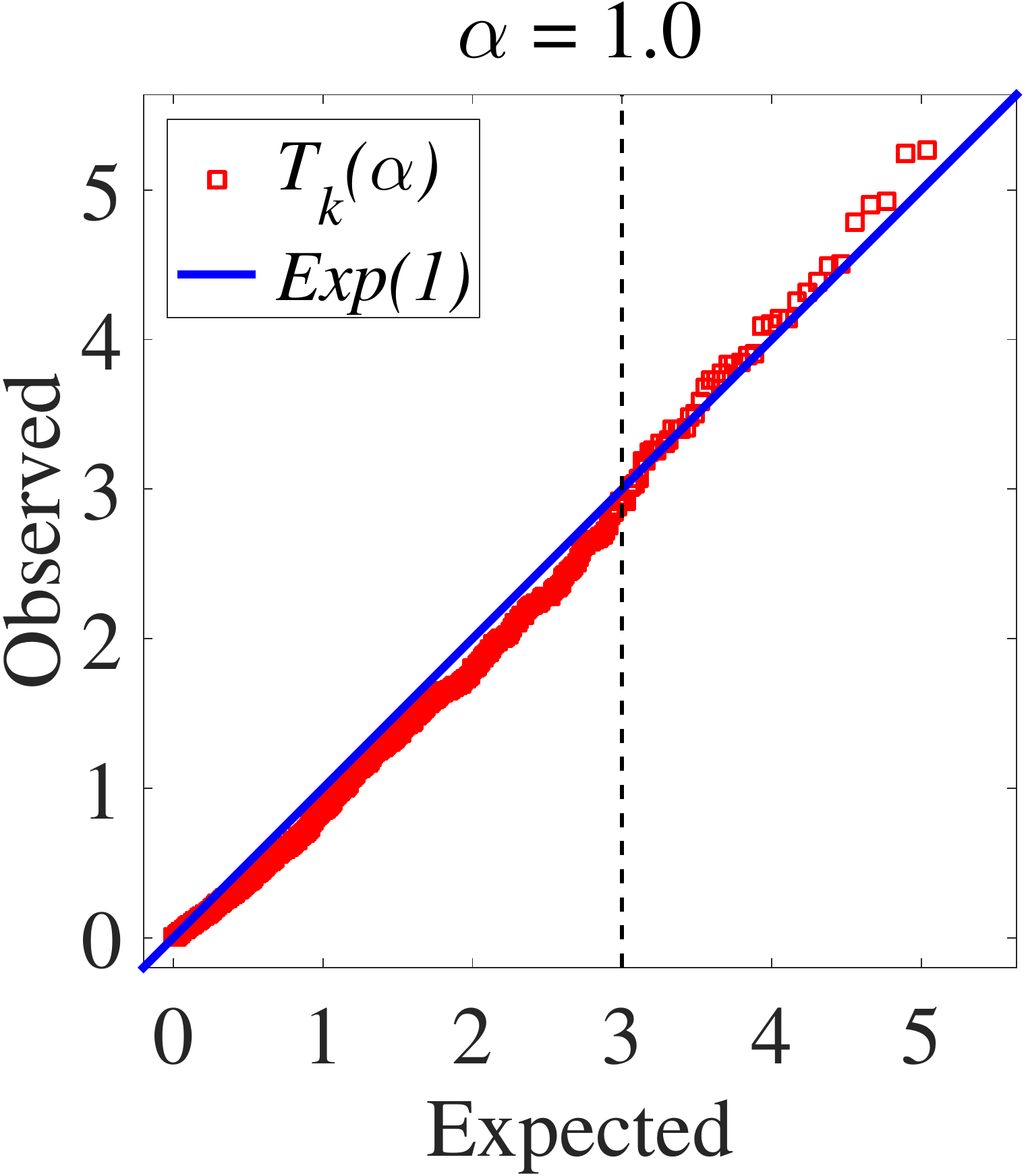}} \;
	\subfloat[]{\includegraphics[width=0.15\textwidth]{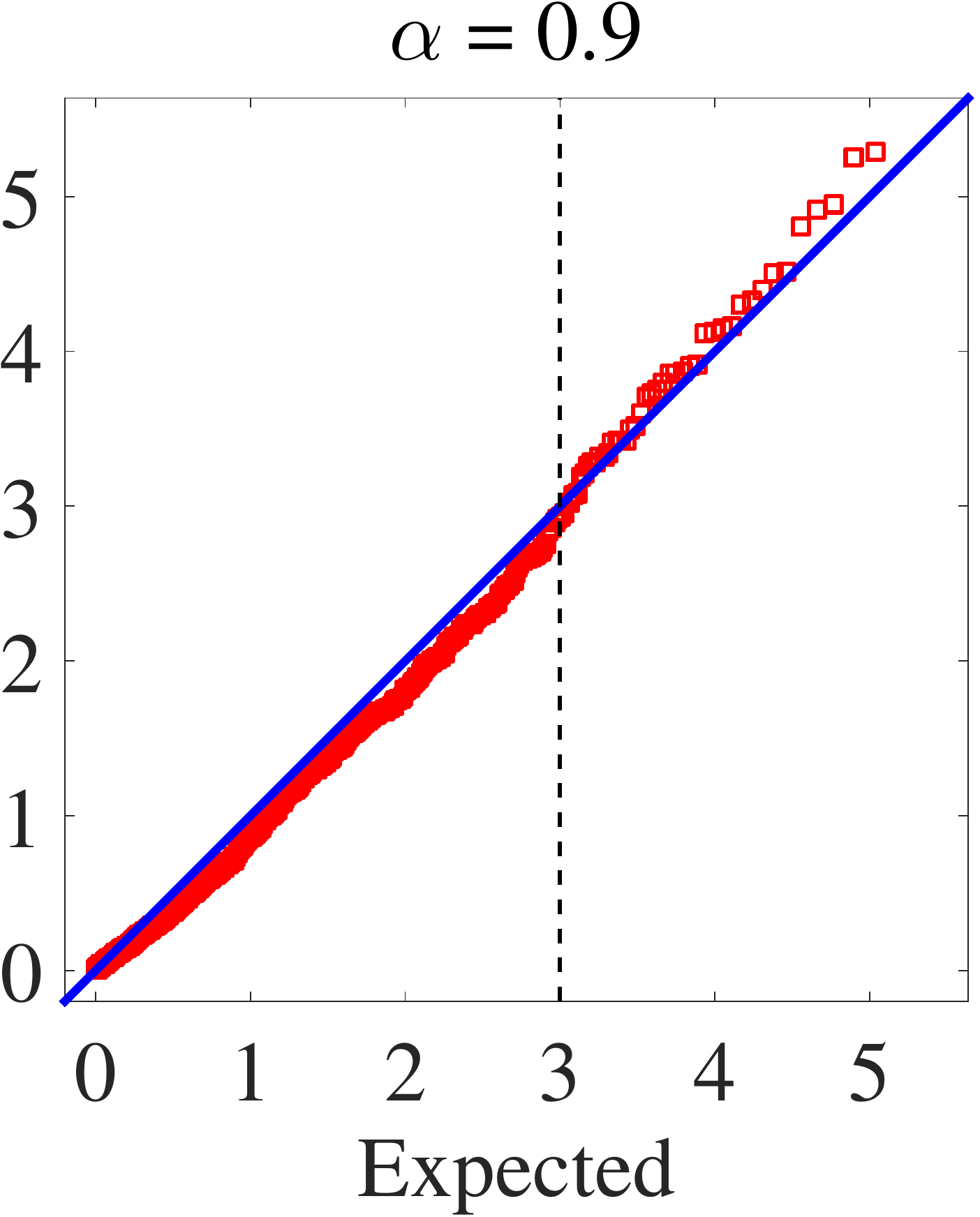}} \;
	\subfloat[]{\includegraphics[width=0.15\textwidth]{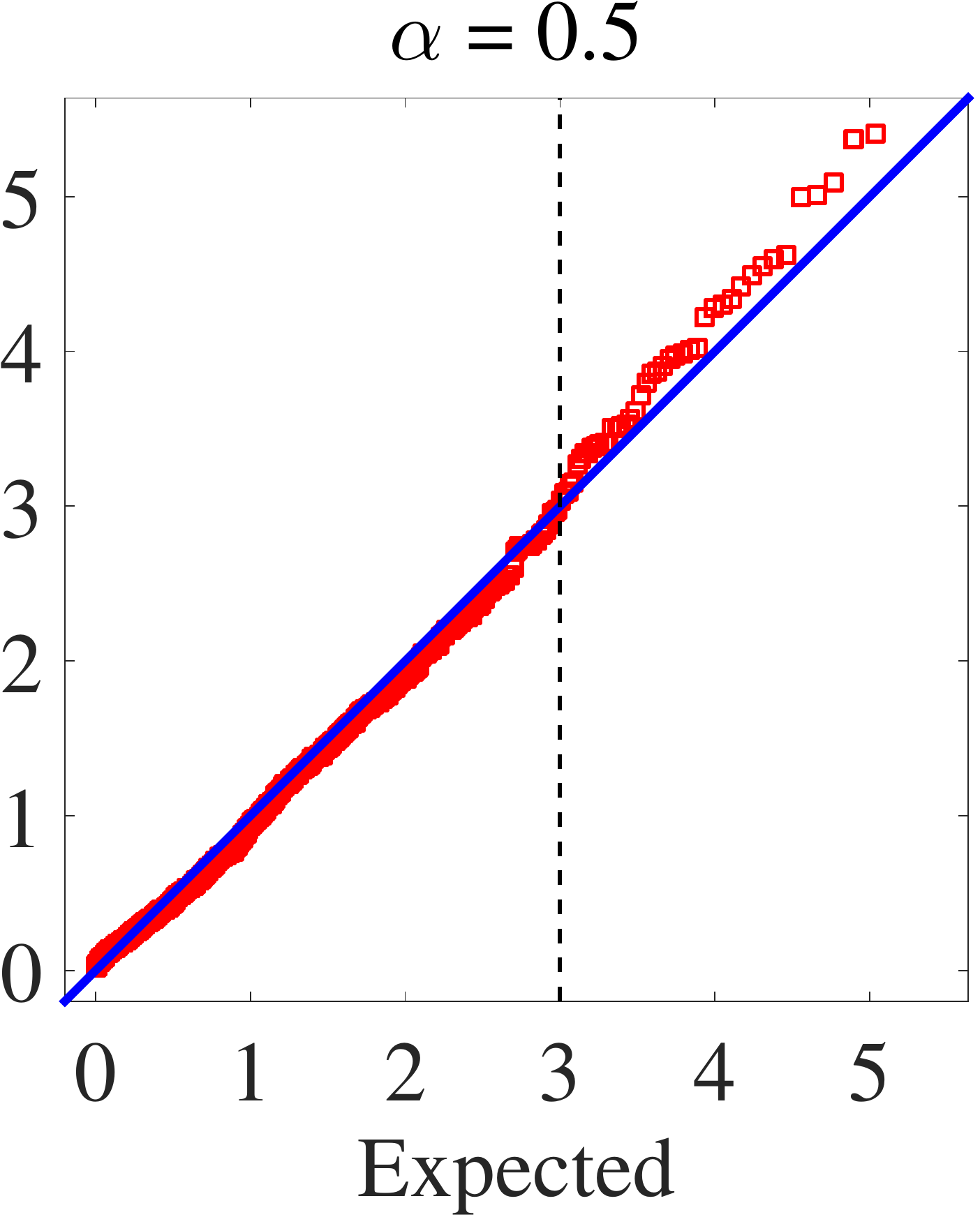}}\hfill

	\subfloat[]{\includegraphics[width=0.163\textwidth]{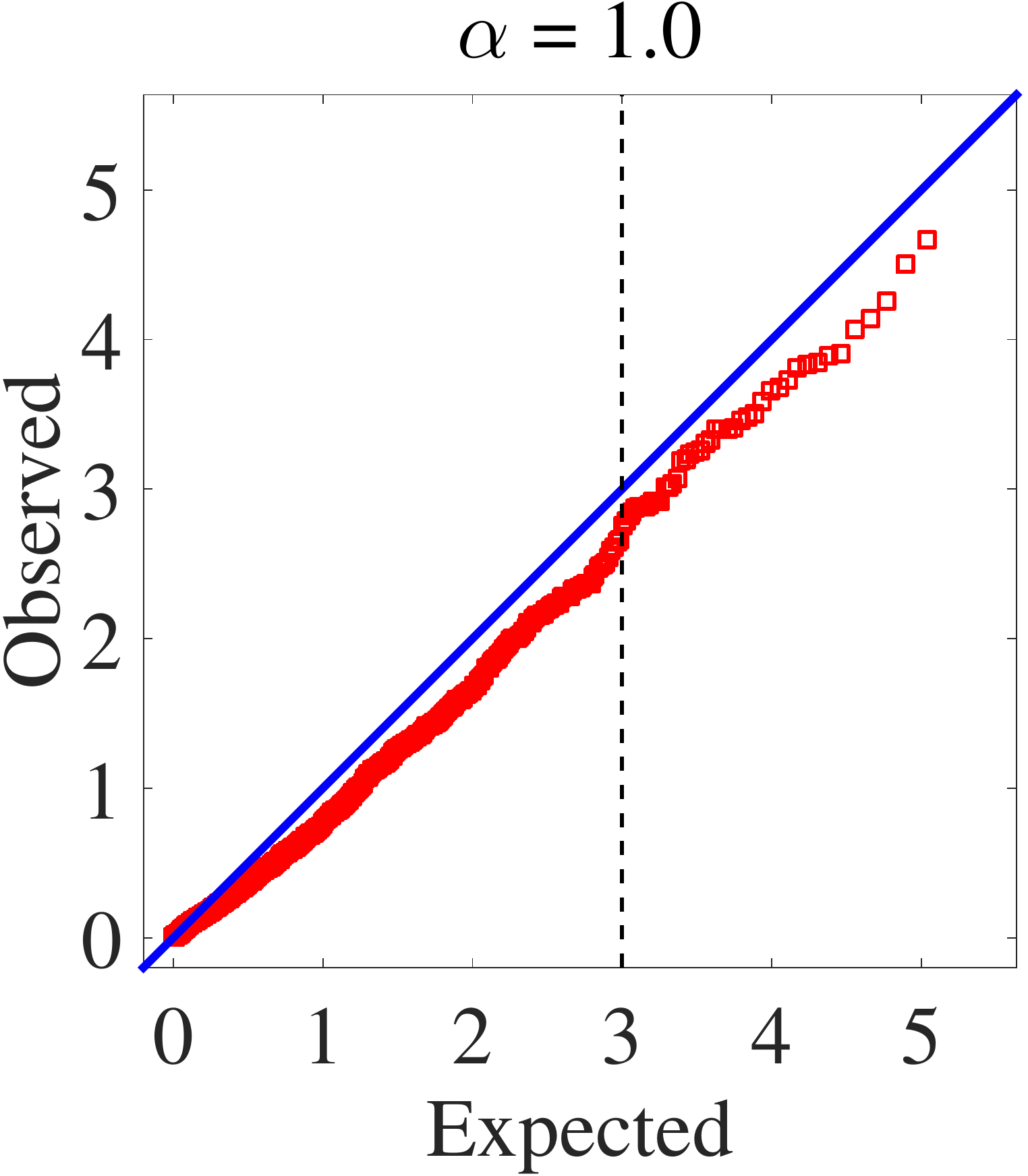}} \;
	\subfloat[]{\includegraphics[width=0.15\textwidth]{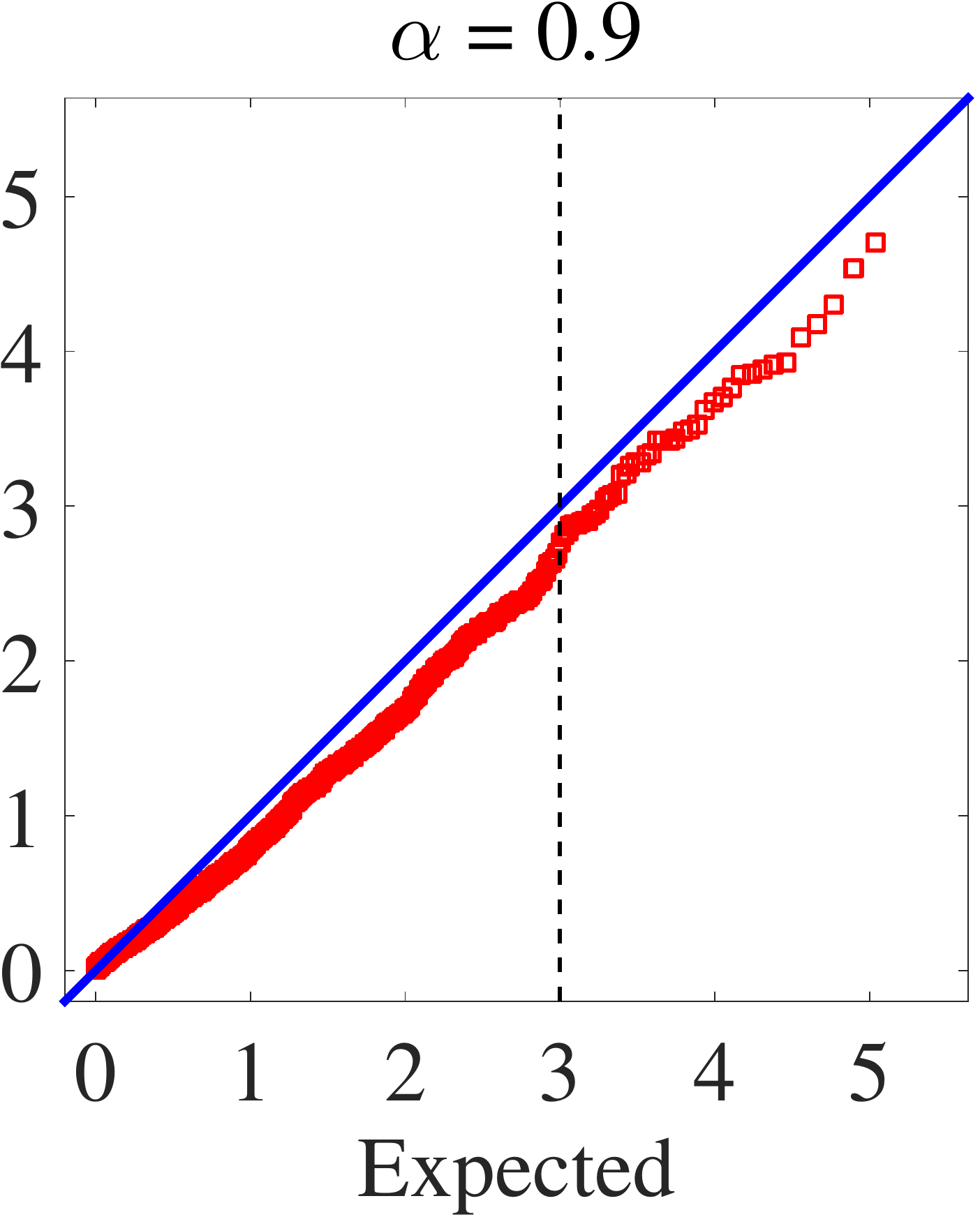}} \;
	\subfloat[]{\includegraphics[width=0.15\textwidth]{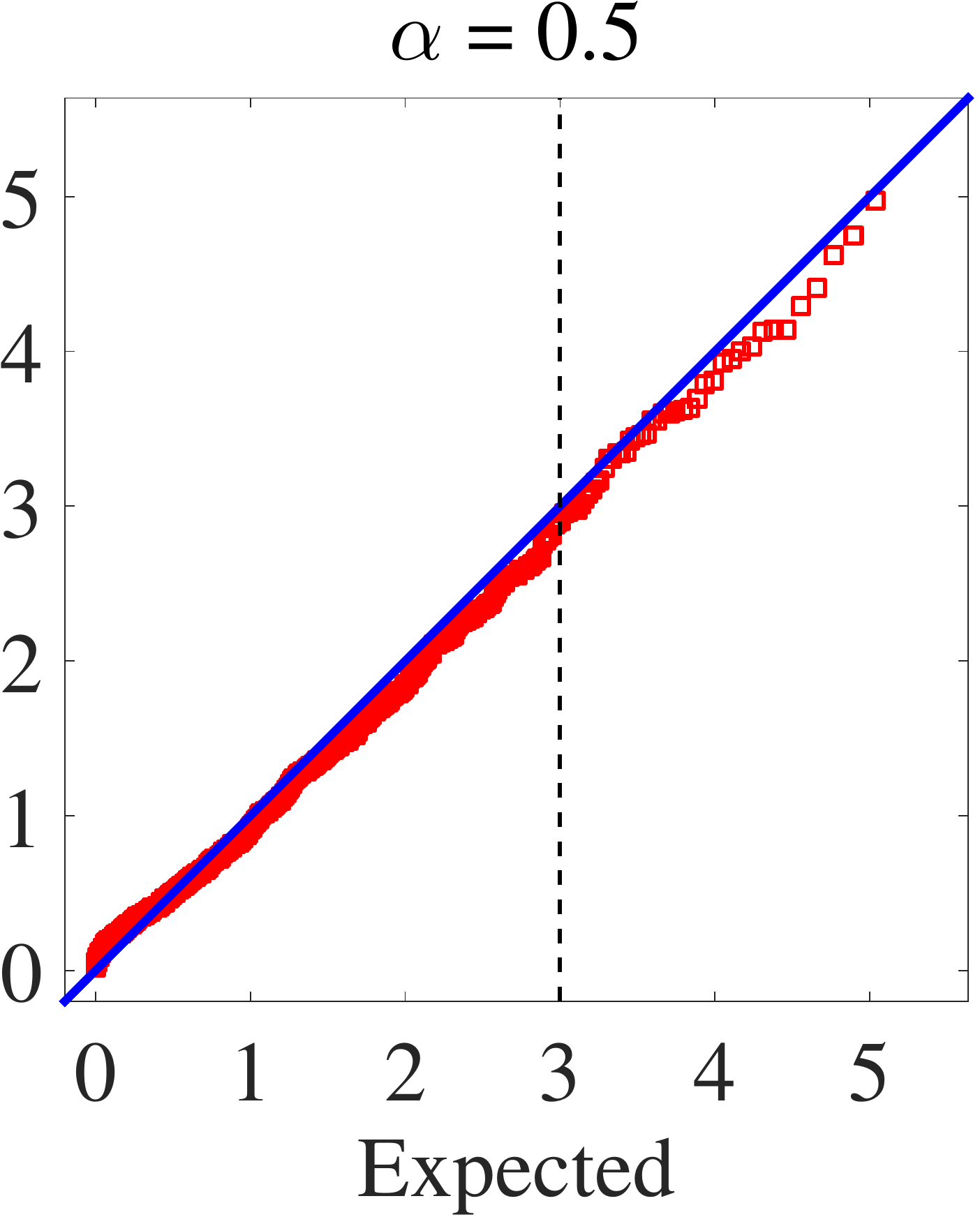}}\hfill
\caption{Quantile-quantile plots, constructed over 1000 simulations, of the EN covariance test statistic $T_{k}(\al)$ in the model, where 
$k = \| \be^* \|_0 \in \{1,2\}$ and $\al  \in \{ 1, 0.9, 0.5\}$.
The first panel gives the results for $k=1$ and the second panel for   $k =2$. The sample size and dimension is $\ndim=100$, $\pdim=50$, respectively.}
\label{fig:TkPlotsD1}
\end{figure}

Let the observations vector $\y \sim \mathcal N_\ndim (\X \bet , \bo I)$, 
the covariance structure for $\x_i \in \R^\pdim$  conforms with $\Sig_1 = \bo I_\pdim$ 
and the number of true non-zero coefficients is varying,  $\| \bet \|_0= k = \{1,2 \}$.  The sample size is $\ndim = 100$ and number of predictors is $\pdim = 50$.  
Note also that the non-zeros values of  $\bet$ are equal to $3$, which are greater than $ \sigma \sqrt{2 \log p}$ as  suggested in \cite[Theorem 1]{lockhart2014CovTest} for $k>1$ case.  
We compute the EN covariance test statistic $T_k(\al)$ for testing the significance of the predictor that enters to the active set $\setA_{k+1}$. 
Fig.~\ref{fig:TkPlotsD1} depicts the quantile-quantile (QQ) plots of the EN covariance test statistic $T_{k}(\al)$, constructed over 1000 MC simulations, for testing the entry (significance) of the $(k+1)^{th}$ predictor, i.e., the predictor that enter the active set in the interval $(\lam_{k},\lam_{k+1})$. The first panel considers the case when $k=1$ and the second panel the case when there are two non-zero predictors, i.e.,  $k =2$. 
As can be seen, the QQ plots illustrate that the test statistic $T_{k}(\al)$ is  well approximated by $\mathrm{Exp}(1)$ distribution. 

\setlength{\textfloatsep}{10pt plus 1.0pt minus 2.0pt}
\graphicspath{ {Results/Figs_Orth/} }

\begin{figure}[!h]
\centering
	\subfloat[]{\includegraphics[width=0.163\textwidth]{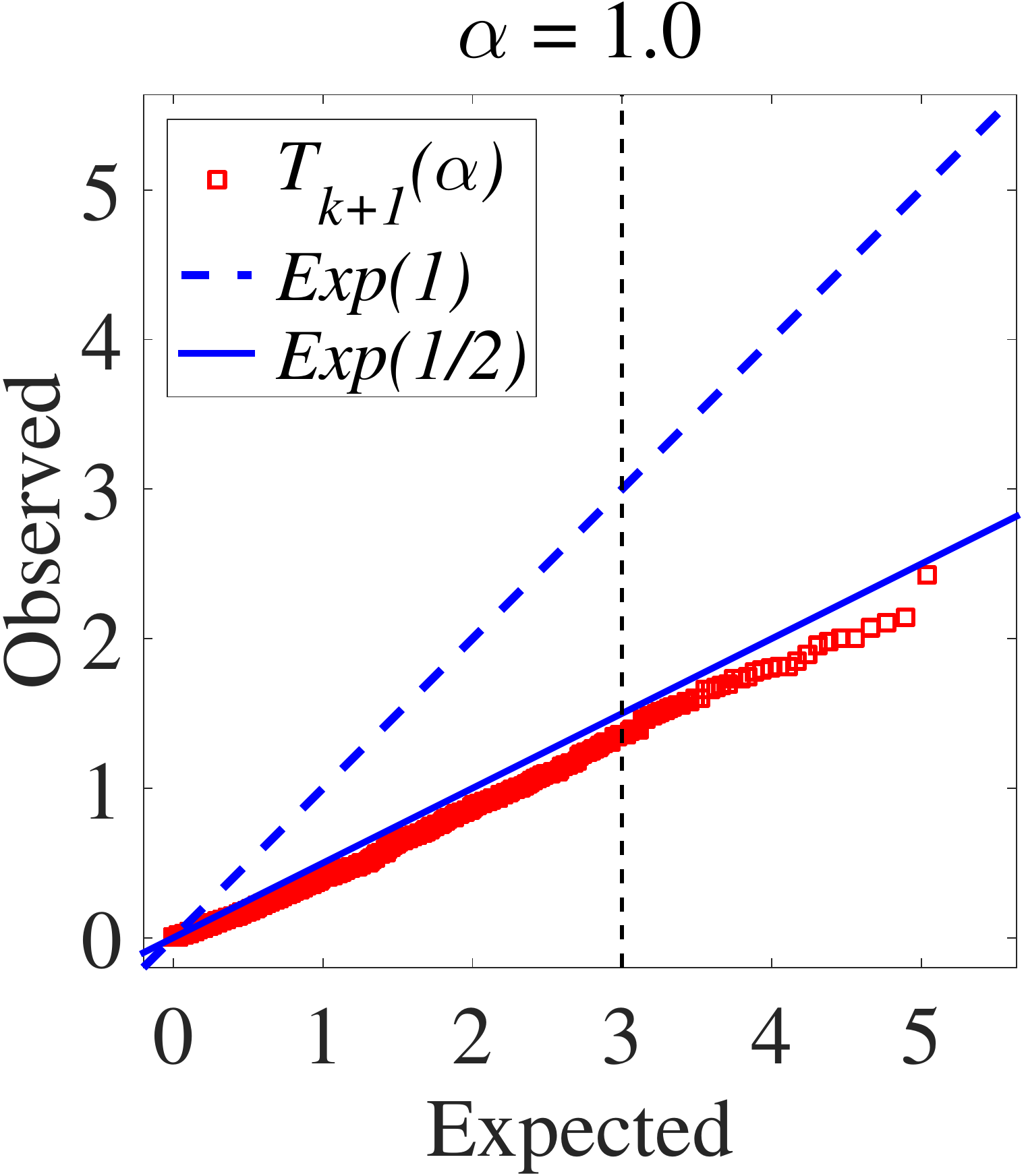}} \;
	\subfloat[]{\includegraphics[width=0.15\textwidth]{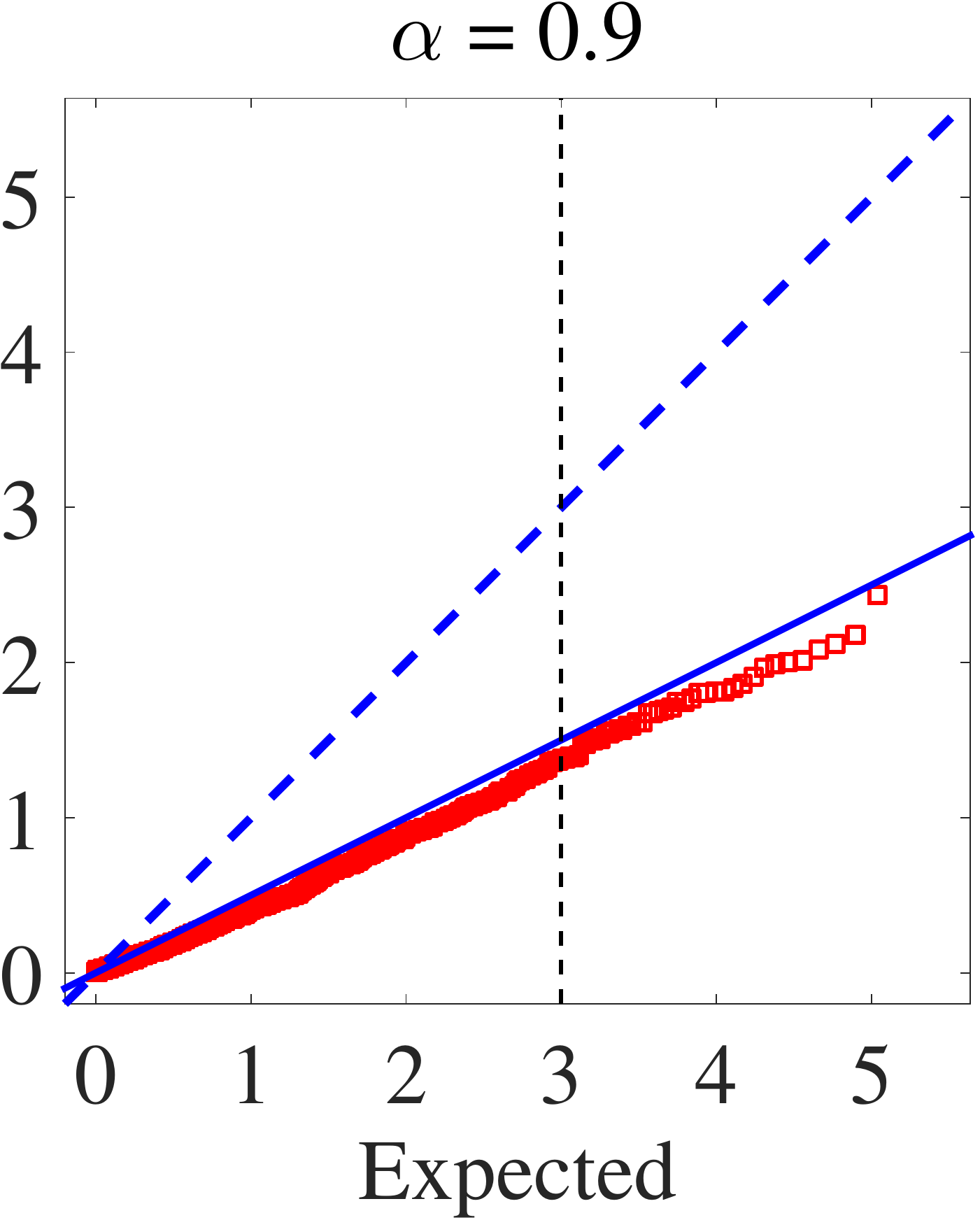}} \;
	\subfloat[]{\includegraphics[width=0.15\textwidth]{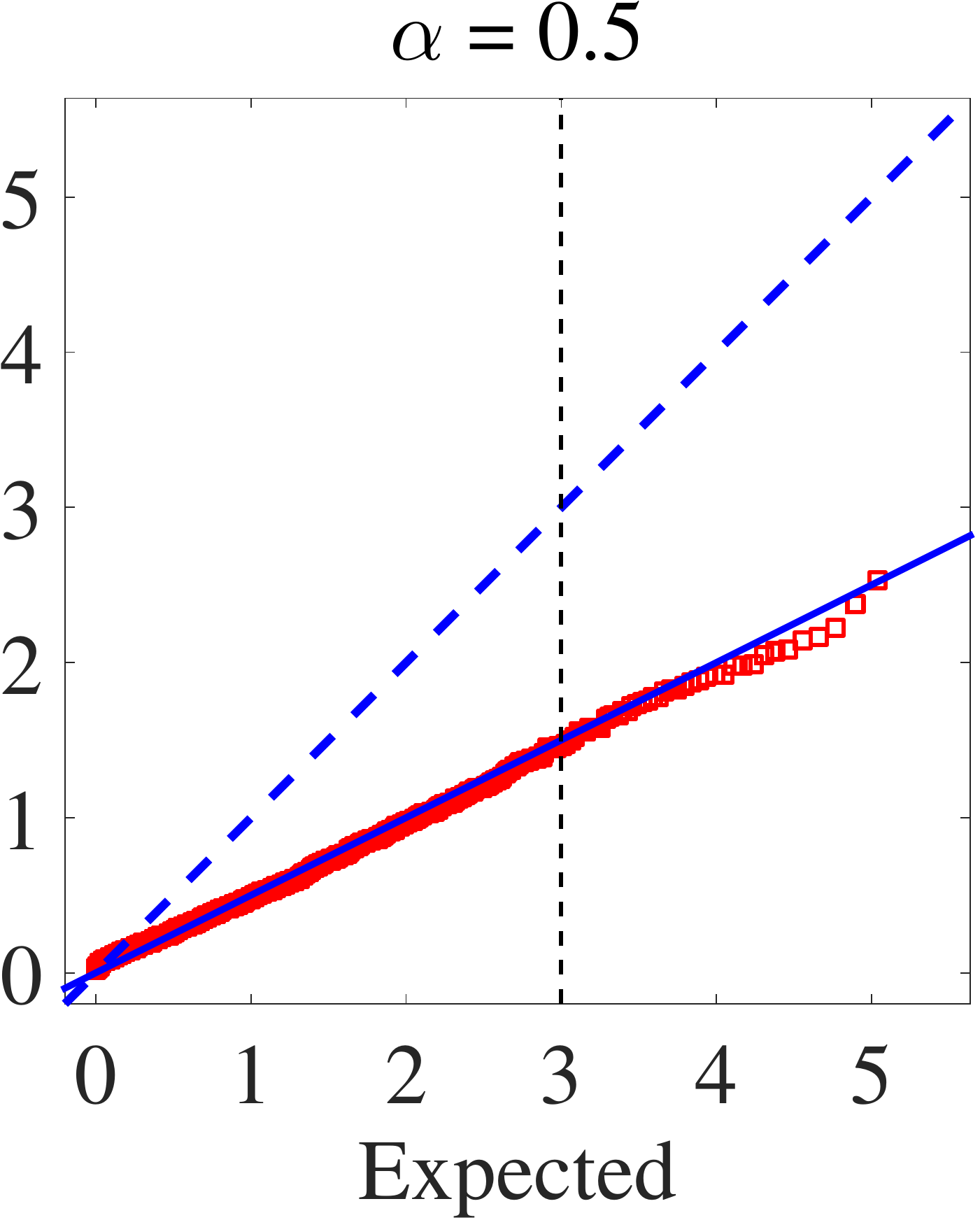}}\hfill
	
\caption{Quantile-quantile plots, constructed over 1000 simulations, of the EN covariance test statistic $T_{k+1}(\al)$ in the model, where 
$k = \| \be^* \|_0 = 1$ and $\al  \in \{ 1, 0.9, 0.5\}$  and $\X^\top \X = \bo I$. 
The sample size and dimension is $\ndim=100$, $\pdim=50$, respectively.}
\label{fig:TkPlotsD2}
\end{figure}

It has been shown in \cite{lockhart2014CovTest} for the case of Lasso and for orthonormal predictors ($\X^\top \X = \bo I$) that  
\beq \label{eq:EN_Tk_d}
\big( T_{k}, \dots, T_{k+d-1}\big) \xrightarrow{\text{\hspace{2pt} d }} \big( \mathrm{Exp}(1),  \dots, \mathrm{Exp}(1/d) \big)
\eeq 
as $\ndim, \pdim \to \infty$.  It is then of interest to investigate if the  EN covariance test statistic  $T_{k}(\al)$ admits the property \eqref{eq:EN_Tk_d} as well but for 
$\al \neq 1$.   We test this property for $T_k(\al)$ in the case that $k = 1$. The simulation set-up is as before except that it has now an orthonormal predictors matrix.
The corresponding QQ plots for different $\al$-values are shown in Fig. \ref{fig:TkPlotsD2}, which clearly illustrate that EN covariance test statistic  $T_{k+1}(\al)$ is well approximated by $\mathrm{Exp}(1/2)$ distribution 
as the result \eqref{eq:EN_Tk_d} suggests.  

\graphicspath{ {Results/Figs_Orth/} }

\section{Real Data Example }\label{sec:data} 
 
A common case in practice is that  $\sigma^2$ is unknown and needs to be estimated from the data. If $\ndim > \pdim$,  one can estimate $\sigma^2$ via 
 \[
\hat \sigma^2 = \|\y - \X \beh_{\mbox{{\tiny LS}}}\|_2^2 / (n-p),
\] 
where  $\beh_{\mbox{{\tiny LS}}}$ is the least-squares estimate for the full model. When one replaces $\sigma^2$ with $\hat \sigma^2$ in covariance test statistic $T_k$  
in \eqref{eq:Tk_lasso} for the Lasso,  its asymptotic distribution changes from  $\mathrm{Exp}(1)$ to $F_{2,\ndim-\pdim}$; see  \cite[Sec. 6]{lockhart2014CovTest} for details. Since we observed in the previous section that for known $\sig^2$, $T_k(\al)$ shares the same asymptotic distribution as $T_k$, it is safe to presume  here that  $F_{2,\ndim-\pdim}$ is an asymptotic distribution of $T_k(\al)$ as well. 

We consider the prostate cancer data set used e.g., in \cite{ESLbook:2001} as well as in \cite[Sec. 6]{lockhart2014CovTest}.
The training data consists of $\ndim=67$ observations of male patients who had surgery for prostate cancer and $\pdim=8$ predictors,   
which are clinical measures that are labelled as 1-'lcavol', 2-'lweight', 3-'age', 4-'lbph', 5-'svi', 6-'lcp', 7-'gleason', 8-'pgg45'. The response is the logarithm of PSA (prostate specific antigen) level.  A more detailed description of the data set can be found in \cite{stamey1989prostate}.  Note that rather strong correlations exist between the predictors: correlation between gleason and  pgg45 is .757,  but high correlations are found between other covariates as well. Correlation  between  svi and lcp  is .673  and .675 between lcavol and lcp.  The condition number of 243.30 which indicates a reasonably strong collinearity in the covariates and  hence EN can be recommended for this data set. 

We calculate the p-values of the EN covariance test statistic $T_k(\al)$ using the $F$-distribution. For the data above, there are $K=\pdim=8$ knots, i.e., at each  active set $\setA_{k}(\al)$, $k=1, \ldots, 8$,  a new variable enters the model. Note that depending on the value of $\al$, the predictors may enter the model in a different order. 
Table \ref{table:t2} lists the p-values along with entering predictor number (in brackets) for all steps. The results are shown for different $\al$-values.
Note that variables enter to the model in different order  depending on the  value of $\al$.  This feature can probably be attributed to existing strong correlations among the predictors. 

\setlength{\textfloatsep}{20.0pt plus 2.0pt minus 4.0pt}
\begin{table}[!t]
\setlength\extrarowheight{2pt}
\caption[caption]{EN Covariance Test Applied to the Prostate Cancer Data Example. The p-values with Entering Predictor Number (in Brackets) for All Steps. \\(The p-values are Rounded to 3 Decimal Places)}
\label{table:t2}
\centering
\begin{tabular}{m{0.4cm} m{0.1cm} m{1.4cm} m{1.4cm} m{1.4cm} m{1.4cm}}
\hline
Step &&$\al=1.0$ &$\al=0.9$ &$\al=0.5$ &$\al=0.1$ \\
\hline\hline
1 &&0.000 (1) &0.000 (1) &0.000 (1) &0.000 (1) \\
2 &&0.052 (2) &0.464 (5) &0.003 (5) &0.000 (5) \\
3 &&0.174 (5) &0.001 (2) &0.053 (2) &0.020 (6) \\
4 &&0.930 (4) &0.470 (8) &0.077 (6) &0.000 (2) \\
5 &&0.353 (8) &0.047 (4) &0.005 (8) &0.000 (8) \\
6 &&0.650 (3) &0.646 (3) &0.011 (4) &0.000 (7) \\
7 &&0.051 (6) &0.055 (6) &0.020 (7) &0.000 (4) \\
8 &&0.978 (7) &0.978 (7) &0.978 (3) &0.167 (2) \\
\hline
\end{tabular}
\end{table}

\section{Conclusions} \label{sec:concl} 

The EN coefficent path for fixed $\al$ is piece-wise linear as a function of the penalty parameter $\lam$. 
In this paper, we proposed a novel approach, pathwise (PW-)LARS-EN algorithm that computes the knots of EN over a grid of $\al$ values. 
Thus the  PW-LARS-EN allows to compute the whole EN solution path over a large range of $\al$ values. Furthermore, we illustrated that the found EN knots  
can be used to construct a significance test for the EN as was done recently in the seminal paper by \cite{lockhart2014CovTest}. 
\section*{Acknowledgment}

The research was partially supported by the Academy of Finland grant no. 298118 which is gratefully acknowledged.



%
\bibliographystyle{IEEEtran}
\bibliography{EUSIPCO2017Ref}

\end{document}